# Atom-Interferometry Constraints on Dark Energy

Paul Hamilton[1], Matt Jaffe[1], Philipp Haslinger[1], Quinn Simmons[1], Holger Müller[1,2*], and Justin Khoury[3]

[1]Department of Physics, 366 Le Conte Hall MS 7300, University of California, Berkeley, California 94720, USA.

[2]Lawrence Berkeley National Laboratory, One Cyclotron Road, Berkeley, CA, 94720, USA.

[3]Center for Particle Cosmology, Department of Physics and Astronomy, University of Pennsylvania, Philadelphia, PA 19104, USA.

*Correspondence to: hm@berkeley.edu.

**Abstract**: If dark energy — which drives the accelerated expansion of the universe — consists of a light scalar field, it might be detectable as a "fifth force" between normal-matter objects, in potential conflict with precision tests of gravity. Chameleon fields and other theories with screening mechanisms, however, can evade these tests by suppressing the forces in regions of high density, such as the laboratory. Using a cesium matter-wave interferometer near a spherical mass in an ultra-high vacuum chamber, we reduce the screening mechanism by probing the field with individual atoms rather than bulk matter. Thus, we constrain a wide class of dark energy theories, including a range of chameleon and other theories that reproduce the observed cosmic acceleration.

Cosmological observations have now firmly established that the universe is expanding at an accelerating pace, which can be explained by dark energy permeating all of space and accounting for ∼ 70% of the energy density of the universe (*1*). What constitutes dark energy, and why it has its particular density, remain as some of the most pressing open questions in physics. What is clear is that dark energy presents us with a new energy scale, of order meV. It is natural to speculate that new (usually scalar) fields might be associated with that scale that make up all or part of the dark energy density (*2*, *3*). String theory with compactified extra dimensions, for instance, features a plethora of scalar fields, which typically couple directly to matter fields unless protected by a shift symmetry as for axions (*4*, *5*). If the fields are light, this coupling would be observable as a "fifth force", in potential conflict with precision tests of gravity (*6*).

Theories with so-called screening mechanisms, on the other hand, have features that suppress their effects in regions of high density, so that they may couple to matter but nonetheless evade experimental constraints (*7*). One prominent example is the chameleon field, whose mass depends on the ambient matter density (*8, 9*). It is light and mediates a long-range force in sparse environments, such as the cosmos, but becomes massive and thus short-ranged in a high-density environment, such as the laboratory (see Fig. **S1**). This makes it difficult to detect by fifth-force experiments.

Burrage, Copeland and Hinds (*10*) have recently proposed to use atom interferometers (*11, 12*) to search for chameleons. An ultrahigh-vacuum chamber containing atomic test particles simulates the low-density conditions of empty space, liberating the chameleon field to become long-ranged and, thus, measurable. Here, we use a cavity-based atom interferometer (*13, 14*) measuring the force between cesium-133 atoms and an aluminum sphere to search for a range of screened dark energy theories that can reproduce the current dark energy density (Fig. **1A, B**).

The chameleon dark energy field $\phi$ in equilibrium is determined by minimizing a potential density $V(\phi)+V_{\text{int}}$, which is the sum of a self-interaction term $V(\phi)$ and a term $V_{\text{int}}$ describing the interaction with ordinary matter. The simplest chameleon theories are characterized by two parameters, having the dimension of mass. The first one, $\varLambda$, enters the self-interaction potential (*15, 16*),

$$V(\phi) = \Lambda^4 e^{\Lambda^n/\phi^n} \simeq \Lambda^4 + \frac{\Lambda^{4+n}}{\phi^n} + \ldots \qquad (1)$$

The term proportional to $1/\phi^n$, where $n$ is a real exponent often taken to be $n=1$, leads to screening while the constant term is responsible for the chameleon's energy density in otherwise empty space. It can drive cosmic acceleration today if $\varLambda = \varLambda_0 \sim 2.4$ meV, given by the current density of dark energy of $7\times10^{-27}$ kg/m$^3$; roughly the mass of four hydrogen atoms per cubic meter. The second parameter, $M$, enters the interaction with ordinary matter of density $\rho$ (again using natural units)

$$V_{\text{int}} = \frac{\phi\rho}{M}. \qquad (2)$$

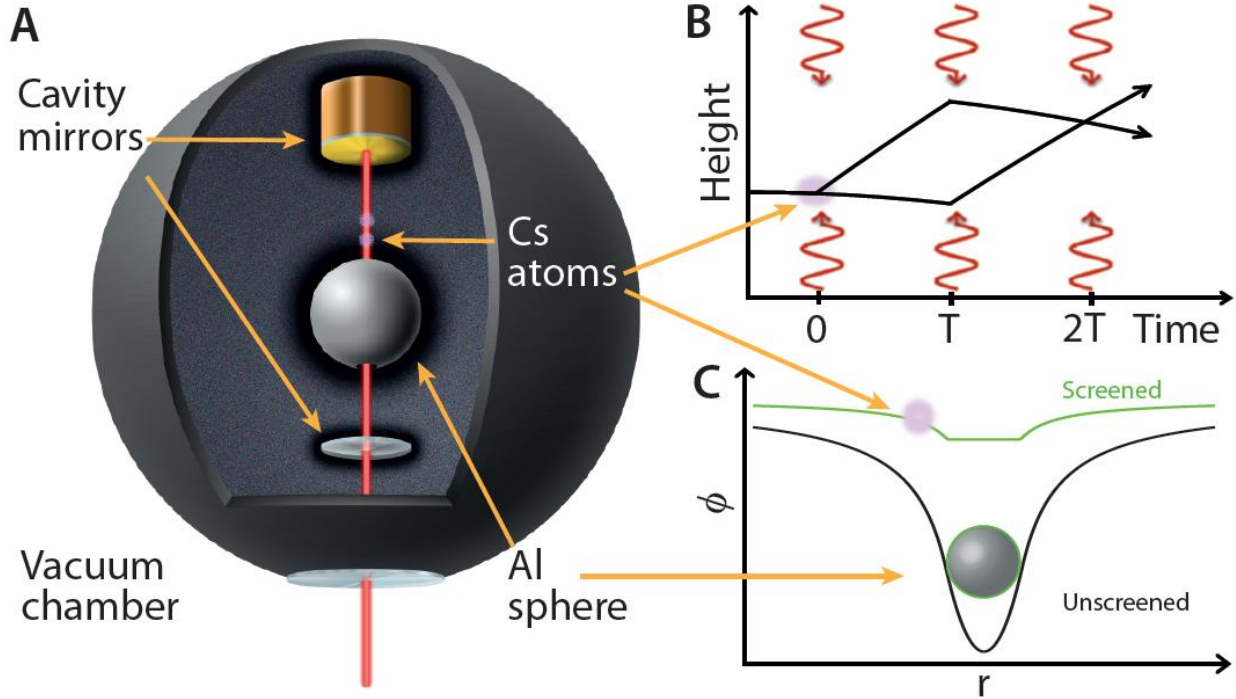

**Fig. 1**. **Screened fields in our experiment** (**A**) The vacuum chamber (radius 5 cm, pressure ~6× 10⁻¹⁰ Torr, mostly hydrogen) holds a pair of mirrors forming a Fabry-Perot cavity and the aluminum (Al) source sphere. Laser beams pass a 1.5-mm radius hole in the $r_s$ = 9.5-mm radius sphere. A Mach-Zehnder interferometer is formed using cold cesium atoms at an effective distance of 8.8 mm from the sphere surface from a magneto-optical trap (not shown). (**B**): Photons in three flashes of laser radiation resonant in the cavity impart momentum to the atoms, directing each atomic matter wave on two paths. (**C**) Potential generated by a macroscopic sphere as function of distance from the center.

The parameter $M$ is essentially unconstrained, but plausibly below the reduced Planck mass $M_{Pl} = (\hbar c/8\pi G)^{1/2} \sim 2.4\times10^{18}$ GeV. A lower bound $M > 10^4$ GeV was derived from hydrogen spectroscopy (*17*).

Existing experimental bounds for $M < M_{Pl}$ come from oscillations of rubidium atoms in a harmonic trap (*18*) and ultracold neutrons (*19, 20*). Limits from astrophysical observations (*7*) and torsion balances (*6, 21*) are available for $M \approx M_{Pl}$, where the chameleon is unscreened. Experiments such as the CHameleon Afterglow SEarch CHASE (*22*), the Axion Dark Matter

eXperiment ADMX (*23*), and the CERN Axion Solar Telescope CAST (*24*) place bounds given an additional coupling of the chameleon to the photon. Our limits do not depend on such extra couplings.

The acceleration of an atom at a radius *r* from the center of the sphere (Fig. **1A**) caused by the sphere via the chameleon interaction and gravity is given by (*10*)

$$a \approx \frac{Gm_s}{r^2}\left[1 + 2\lambda_a\lambda_s\left(\frac{M_{\text{Pl}}}{M}\right)^2\right], \tag{3}$$

where *G* is the gravitational constant and $m_s$ the mass of the sphere. The screening factors $\lambda_a$ and $\lambda_s$ for the atom and the sphere, respectively, are functions of the object's mass and radius as well as the parameters $\Lambda$, *M,* and *n* of the theory. See Eq. (S1). They approach 1 for small and light particles. For macroscopic objects, however, only a thin, outermost layer will interact with the chameleon field (Fig. **1C**), leading to a screening factor much smaller than one. Macroscopic fifth-force experiments are faced with two small screening factors but atom interferometers avoid this double suppression.

The operation of the atom interferometer is based on the matter-wave concept of quantum mechanics. When the atom absorbs or emits a photon, it recoils with the momentum $\hbar k$ (where $\hbar$ is the reduced Planck constant and *k* the wavenumber of the photon). We use a two-photon Raman transition between the two hyperfine levels of the ground state of cesium, which are labeled by their total angular momentum quantum numbers of *F* = 3 and 4, respectively. The transition is driven by two vertical, counterpropagating laser beams, see Fig. **1A**. The atom absorbs a photon from the first beam and is stimulated by the second beam to emit a photon into the opposite direction. The net effect on the atom is a change of the internal quantum state from *F* = 3 to *F* = 4 and an impulse of $\hbar k_{\text{eff}}$, where the effective wavenumber $k_{\text{eff}}$ is the sum of the wavenumbers of the two beams. The duration and intensity of the laser pulses can be tuned such that the transfer happens with 50% probability or nearly 100%, forming beam splitters and mirrors, respectively, for matter waves.

Our Mach-Zehnder interferometer (Figure **1B**) uses a sequence of three light pulses separated by equal time intervals *T*. The first pulse splits the matter-wave packet describing each atom into two partial ones that separate with a recoil velocity of about 7 mm/s. The second acts as a mirror that reverses the direction of the relative motion, and the third is a beam splitter that overlaps the partial wave packets. Interference of the partial matter waves determines the probability $P$ of the atoms to arrive in each of the two interferometer outputs,

$$P = \cos^2(\Delta\phi/2), \tag{4}$$

where the phase difference accumulated between the partial wave packets (*11*)

$$\Delta\phi = k_{eff} a_{tot} T^2 \tag{5}$$

is a function of the total acceleration $a_{tot}=a+g$ of the atoms, the sum of the acceleration due to chameleon-mediated interactions with the sphere, Eq. (3), and the far larger acceleration $g$ due to Earth's gravity (and small systematic effects).

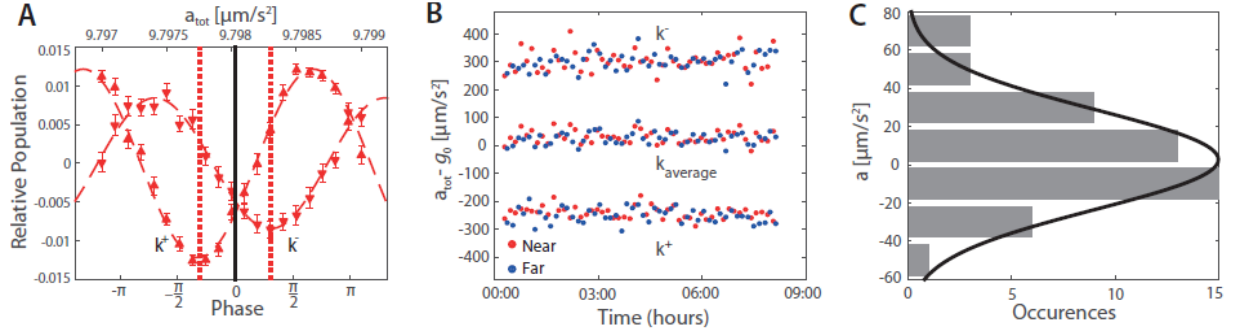

**Fig. 2. Data.** (**A**): Two interference fringes, measured with the wavevector up (k+) and inverted (k-). (**B**) Acceleration $a_{tot}-g_0$, where $g_0 = 9.798$ m/s², measured with wavevector-reversal and sphere near (red) and far (blue). The top group of data has the wavevector pointing downward, the bottom group upward. The plotted data is from a total of 16,800 runs. Taking the average (middle) suppresses systematic effects. Data taken during the night before about 6:30 shows lowest noise, suggesting that our sensitivity is limited by vibrations. (**C**) Histogram of differences between subsequent measurements with the sphere in the near and far positions.

The most sensitive atom interferometers use pulse separation times $T \sim 1$ s, over which the atoms fall up to $\sim 10$ m in tall atomic fountains (*25-27*). We, however, must keep the atoms within a few millimeters of the sphere to sample the highest chameleon field gradient, and are thus constrained to $T \sim 10$ ms, resulting in a ten thousand fold signal reduction. Our cavity-based atom interferometer (*12, 28*), however, reaches relatively high resolution under these constraints.

A full experimental run takes 1.7 seconds. We prepare about 10 million cesium atoms at a temperature of 5 microkelvin in the $F = 3$ state, using a two-dimensional magneto-optical trap (2D-MOT) to load a 3D-MOT through a differential pumping stage. We run the interferometer with a pulse separation time of $T = 15.5$ ms and detect the two outputs separately using fluorescence detection with a camera (*14*).

Figure **2A** shows an interference fringe obtained by measuring the atom number at the two interferometer outputs while varying the phase $\Delta\phi$ (*13, 14*). Fitting the fringe with a sinewave determines the total acceleration of the atoms. To take out systematic effects, we apply wavevector-reversal, i.e., change the direction of the photon impulse. This inverts the signal due to accelerations but many systematic effects remain unchanged and can be taken out (*29*). To measure the acceleration $a$ originating from atom-sphere interactions (our signal for chameleons) separately from Earth's gravitational acceleration $g$, we compare the total acceleration $a_{tot}=a+g$ with the sphere located in the "near" position to $g$, measured with the sphere in the "far" position. "Near" means an effective vertical distance of 8.8 mm from the surface of the sphere, and "far" means about 3 cm to the side.

One measurement consists of four interference fringes, one each with the wavevector normal and inverted, with the sphere near and far. Figure **2B** shows 50 such measurements with their statistical error bars. For each, we average the acceleration as measured with normal and inverted wavevector to eliminate systematic effects, and compare the acceleration thus measured between the sphere near and far. Figure **2C** shows a histogram of these acceleration differences. Fitting a Gaussian to the histogram results in an estimate of $a = (2.7 \pm 3.3)$ µm/s$^2$. We add corrections for systematic ac Stark effects, magnetic fields, and electrostatic fields (*13*), Table **S1**, and arrive at $a = (-0.7 \pm 3.7)$ µm/s$^2$. The negative sign indicates acceleration away from the sphere. The two-sigma (95%) confidence interval for this data is $-8.2$ µm/s$^2 < a < +6.8$ µm/s$^2$.

A chameleon has spin 0 and can therefore only produce attractive forces (assuming universal coupling to matter). A one-tailed test shows $a < 5.5$ µm/s$^2$ at 95% confidence level. Comparison to the expected acceleration, Eq. (S8-S11) yields the excluded range of parameters $\Lambda$ and $M$ shown in Figure **3A**. Our experiments excludes chameleons at the scale of the cosmological constant $\Lambda =\Lambda_0 =2.4$ meV for $M <2.3\times10^{-5}$ $M_{Pl}$, making the most conservative assumption $\xi=0.55$ for a parameter entering Eq. (S9, S10) that describes the influence of the vacuum chamber walls (*14*). This result rules out chameleons that would reproduce the observed acceleration of the cosmos. To place our result in the context of previous experiments, we now assume that $\Lambda=\Lambda_0$. Fig. **3B** shows the excluded region for different values of the exponent $n$, and Fig. **3C** shows the excluded region compared to experiments that assume photon-chameleon coupling; our result does not rely on such a coupling. In short, the only chameleon theories still viable are the white areas in Figs. (**3A-C**), all of which we have narrowed by several orders of magnitude using atom interferometry.

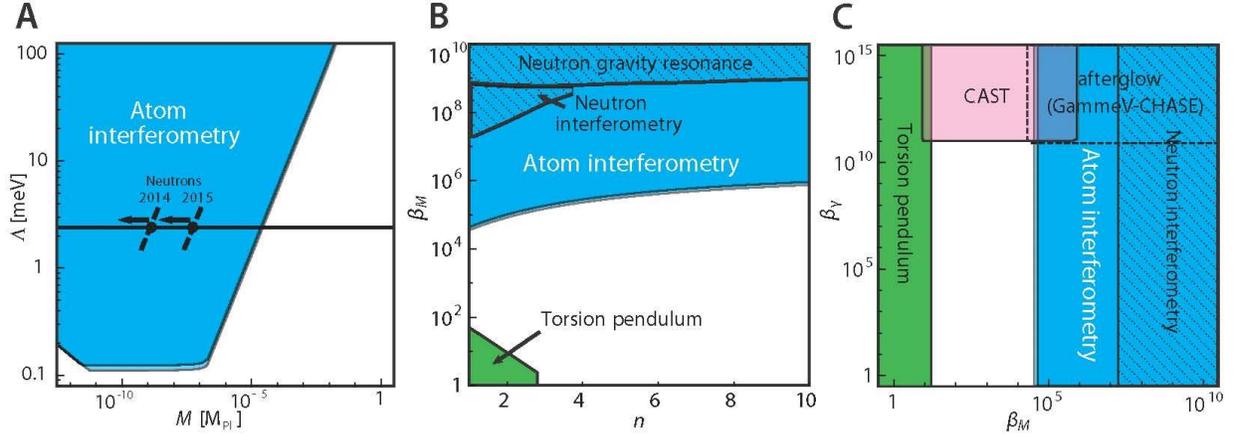

**Fig. 3. Regions of exclusion.** Blue areas are ruled out by our experiment. The narrow light blue stripes at their border show the influence of the variation of $0.55 \leq \xi \leq 0.68$ arising from different models for the boundary of the vacuum chamber (*14*), demonstrating the robustness of our limits. (**A**) Region excluded at 95% confidence level in the $M$-$\Lambda$ plane for $n=1$ in the self-interaction potential Eq. (1). The horizontal line marks the range around $\Lambda_0 = 2.4$ meV where the chameleon field would reproduce the current cosmic acceleration. Also indicated are the highest values of $M$ excluded by neutron experiments (*19, 20*). (**B**) Comparison to neutron gravity resonance (*19*) and neutron interferometry (*20*) in the $n$-$\beta_M$ plane, where $\beta_M = M_{Pl}/M$, assuming $\Lambda=\Lambda_0$. Our results are significantly lower for all values of the exponent $n$ and $\beta_M$. Torsion pendulum experiments (*6, 21*) limit chameleons from the other (low $\beta_M$) end. (**C**) Comparison with CHASE (*22*), ADMX (*23*), and CAST (*24*), experiments that assume photon coupling, assuming $n=1$ and $\Lambda=\Lambda_0$. Atom interferometers as well as neutron and torsion balance experiments give bounds that are independent of the photon coupling parameter $\beta_\gamma$.

The analysis can be generalized to constrain other scalar field theories, such as symmetron, varying-dilaton, and $f(R)$ theories. These theories belong to the same universality class as the chameleon, in that their screening effect is triggered by the local scalar field value as opposed to its spatial derivatives. As a result, their phenomenology is similar to the chameleon (*7*).

**Acknowledgments:** We acknowledge important discussions with Dimitry Budker, Clare Burrage, Andrew Charman, Yasunori Nomura, Saul Perlmutter, Surjeet Rajendran, and Paul Steinhardt. This work was supported by the David and Lucile Packard Foundation, the DARPA Young Faculty Award N66001-12-1-4232, NSF grant Phy-1404566, and NASA grants NNH13ZTT002N, NNH13ZTT002N, and NNH11ZTT001N. P. Haslinger thanks the Austrian Science Fund (FWF): J3680. The work of JK is supported by NSF

CAREER Award PHY-1145525 and NASA ATP grant NNX11AI95G. This paper is dedicated to Martin Perl with gratitude.

**Supplementary Text**

The Chameleon mechanism

Figure **S1** compares the chameleon effective potential $V_{\text{eff}}$ (solid curves) in high-density and low-density environments. It is the sum of the potential $V(\phi) = \Lambda^{4+n}/\phi^n$ (dashed curves) and the linear coupling to matter $\beta_M \phi \rho / M_{\text{Pl}}$ (dotted curves) where $\beta_M = M_{\text{Pl}}/M$. In regions of low density, the minimum of the effective potential lies at large field values $\phi$ and is shallow, corresponding to small-mass chameleon particles, $m^2 = \partial^2 V_{\text{eff}}/\partial \phi^2$. In regions of high density, the minimum lies at small field values and is highly curved, corresponding to high-mass chameleon particles. Thus the mass of chameleons is an increasing function of density, making any chameleon-induced forces short-ranged and therefore screened in typical fifth-force experiments.

Chameleon-Photon coupling

Although not necessary, the chameleon may be coupled to electromagnetism via a term $\exp(\beta_\gamma \phi / M_{\text{Pl}}) F^{\mu\nu} F_{\mu\nu}$ added to the effective potential, where $\beta_\gamma$ describes the strength of this coupling and $F_{\mu\nu}$ is the electromagnetic field strength. In the presence of a magnetic field, this results in chameleon-photon oscillations, akin to axion-photon oscillations (*23*). Our atom interferometry constraints do not rely on such coupling. The chameleon constraints in the $\beta_M$, $\beta_\gamma$ plane are shown in Fig. **3C**.

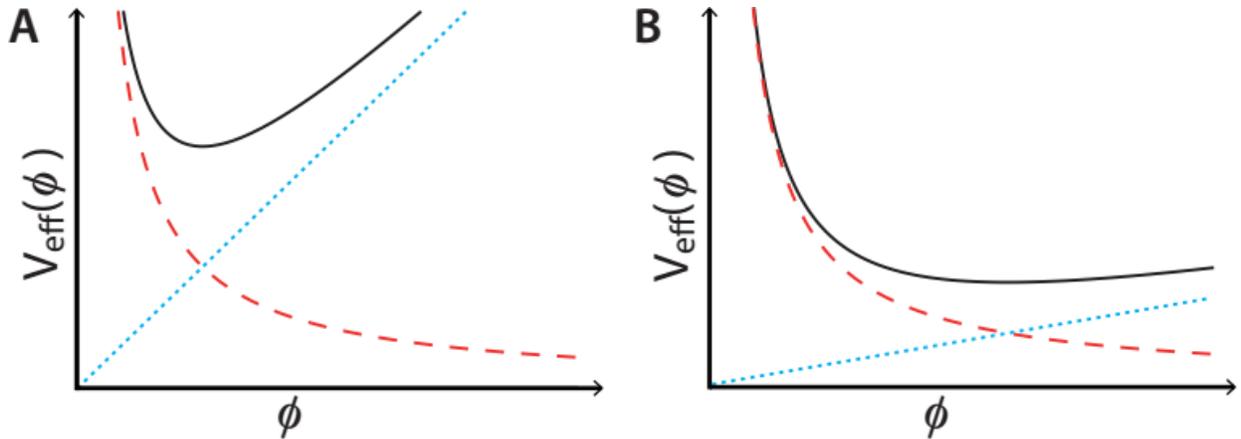

**Fig. S1. Chameleon effective potential versus chameleon field ϕ.** In a high-density environment (A), the effective potential is minimized at a low value of $\phi$, and the curvature in this region is large, leading to a large mass of the chameleon particle and thus a screened (short-ranged) force. In low-density environments, however (B), the field adopts a large equilibrium value where curvature and thus the mass are low.

Calculation of Chameleon-induced forces

We treat Eq. (3) as resulting from a quantum effective potential, as is common (*10*). We assume a universal coupling to matter for simplicity, though our analysis can straightforwardly be generalized to chameleons that couple with different strength to different matter species. To predict the force, we follow (*10*) but generalize to the case of $n\neq 1$ in Eq. (1). We assume a sphere of radius $r_s$, density $\rho_s$, and mass $m_s$, and model the atoms as spheres having the nuclear mass $m_a$, radius $r_a$, and density $\rho_a$. The acceleration of the atom resulting from the chameleon is given by Eq. (3). When an object is screened, the chameleon field couples only to a thin shell of thickness $s_i = r_i\sqrt{1 - 2M\phi_{bg}/(\rho_i r_i^2)}$, where $\phi_{bg}$ is the background field value inside the vacuum chamber. near the surface of the object (Fig. **1C**). This suppresses the force by the screening factors (this section uses natural units $c=\hbar=1$ throughout)

$$\lambda_i \simeq \begin{cases} 1; & \rho_i r_i^2 < 2M\phi_{bg}, \\ 1 - s_i^3/r_i^3; & \rho_i r_i^2 > 2M\phi_{bg}, \end{cases} \tag{S1}$$

where the index $i =$ a, s denotes the atom and the sphere, respectively.

For small $M$ and large density of the test particles, even a microscopic particle may be screened. There are two hypotheses in the literature about this screening. We adopt the conservative one (*10*) that this screening is determined by the high density of the atomic nucleus, which is why our limits in Fig. **3A** level off at $\Lambda \simeq 0.1$ meV. Another hypothesis, using the much lower mass density of the quantum mechanical wave packets, has been used in analyzing previous neutron experiments (*19, 20*). Fortunately, it is unimportant for our purposes to decide which hypothesis is correct, as our limits reach below $\Lambda_0$ at the same value of $M$ regardless (they would, however, continue to lower values of $\Lambda$ if screening was determined by the wavepacket density). Also, this question does not affect the results of (*20*) when they are shown as in Fig. 3B.

The background field value, $\phi_{bg}$, is the smallest of $\phi_{eq}$, the equilibrium value for the residual gas inside the vacuum chamber, and $\phi_{vac}$, which is set by the size of the chamber:

$$\phi_{bg} = \min(\phi_{eq},\ \phi_{vac}). \tag{S2}$$

For a generalized potential Eq. (1), $\phi_{eq}$ is determined by the equilibrium condition

$$-\frac{n\Lambda^{4+n}}{\phi_{eq}^{n+1}} + \frac{\rho_v}{M} = 0, \tag{S3}$$

where $\rho_v$ is the average density of residual gas in the chamber, which is mostly hydrogen. Therefore,

$$\phi_{eq} = \left(\frac{nM\Lambda^{4+n}}{\rho_v}\right)^{\frac{1}{1+n}}. \tag{S4}$$

Meanwhile, the vacuum value $\phi_{vac}$ can be approximated by equalizing the chameleon Compton wavelength and the characteristic size $r_v$ of the vacuum chamber:

$$m_{vac}^2 = \left.\frac{\partial^2 V}{\partial \phi^2}\right|_{vac} = \frac{n(n+1)\Lambda^{4+n}}{\phi_{vac}^{n+2}} \propto \frac{1}{r_v^2}. \tag{S5}$$

In other words,

$$\phi_{vac} = \xi[n(n+1)\Lambda^{4+n}r_v^2]^{1/(n+2)}. \tag{S6}$$

The order-unity factor $\xi$ depends on the geometry of the vacuum chamber and is obtained by comparison with the value of $\phi$ at the center of the vacuum chamber as determined from a numerical solution of the chameleon equation of motion. For example, in polar coordinates,

$$\frac{d^2\phi(r)}{dr^2} + \frac{2}{r}\frac{d\phi(r)}{dr} + n\frac{\Lambda^{4+n}}{\phi(r)^{n+1}} - \frac{\rho(r)}{M} = 0. \tag{S7}$$

This can be numerically integrated straightforwardly, subject to the boundary conditions $f'(0) = 0$ (regularity at the origin) and $\phi \to \phi_{envir}$ (the equilibrium value for environmental density) far from the chamber.

Numerical integration of (S7) shows that the coefficient $\xi$ is largely insensitive to $M$, $n$, and $r$, as well as the chamber geometry. To demonstrate this, we first explore the sensitivity of $\xi$ to the geometry of the vacuum chamber by considering three simple cases: a sphere, an infinite cylinder, and a one-dimensional (1D) plane as used in the neutron analysis (*19, 20*). As shown in Table S1, despite dramatic differences in geometry, $\xi$ only varies between 0.6 (sphere) and 0.8 (1D). Discarding the 1D model as too unrealistic, we take $0.6 \leq \xi \leq 0.68$ as the variation due to the chamber geometry.

**Table S1.** Sensitivity of $\xi$ to chamber geometry, assuming $r_v = 5$ cm, $n = 1$, $\Lambda = 0.1$ meV and $M = 10^{-3} M_{Pl}$, and with boundary condition $\phi \to \phi_{atm}$ as $r \to \infty$.

| Geometry | $[n(n+1)\Lambda^{4+n}r_v^2]^{1/(n+2)}$ | $\phi_{vac}$ | $\xi$ |
|---|---|---|---|
| Sphere | 1.08 meV | 0.66 meV | 0.61 |
| Cylinder | 1.08 meV | 0.73 meV | 0.68 |
| Line | 1.08 meV | 0.89 meV | 0.82 |

Second, we explore the sensitivity of $\xi$ to the assumed boundary conditions for the scalar field, by considering three cases: (i) surrounding the vacuum volume by air so that $\phi(r_v) \to \phi_{air}$, (ii) surrounding the vacuum by material having a density of $\rho = 10$ g/cm$^3$ so that $\phi(r_v) \to \phi_{steel}$, (our actual chamber with ~1-cm thick walls of steel having $\rho \sim 7$ g/cm$^3$ lies between these two cases), and (iii) imposing a boundary condition of $\phi(r_v) = 0$, as would follow from assuming infinitely dense surroundings. For this comparison, we focus on the spherical geometry for concreteness. Table S2 shows that $\xi$ only varies between 0.55 (ii, iii) and 0.61 (i). This analysis also shows the robustness of $\xi$ against variations of the parameter $M$ because $\rho$ and $M$ enter the differential equation (S7) only in the combination $\rho/M$.

**Table S2.** Sensitivity of $\xi$ to asymptotic boundary condition (same parameters as in Table S1), assuming the spherical geometry.

| Geometry | $[n(n+1)\Lambda^{4+n}r_v^2]^{1/(n+2)}$ | $\phi_{vac}$ | $\xi$ |
|---|---|---|---|
| $\phi \to \phi_{atm}$ as $r \to \infty$ | 1.08 meV | 0.66 meV | 0.61 |
| $\phi \to \phi_{steel}$ as $r \to \infty$ | 1.08 meV | 0.59 meV | 0.55 |
| $\phi(r_v) = 0$ | 1.08 meV | 0.59 meV | 0.55 |

**Table S3.** Sensitivity of $\xi$ to power $n$ (same parameters as in Table S1), assuming the spherical geometry and with boundary condition $\varphi \to \varphi_{\text{atm}}$ as $r \to \infty$.

| $n$ | $[n(n+1)\Lambda^{4+n}r_v^2]^{1/(n+2)}$ | $\phi_{\text{vac}}$ | $\xi$ |
|---|---|---|---|
| 1 | 1.08 meV | 0.66 meV | 0.61 |
| 2 | 0.78 meV | 0.43 meV | 0.55 |
| 3 | 0.60 meV | 0.33 meV | 0.56 |
| 4 | 0.48 meV | 0.28 meV | 0.58 |
| 5 | 0.41 meV | 0.24 meV | 0.60 |

Next, Tab S3 shows that variations of the chameleon exponent $n$ between 1 and 5 leave $\xi$ within the range of $0.55 \leq \xi \leq 0.61$, see Tab S3. Finally, Fig. S2 shows a numerical solution of Eq. (S7) as function of $r$ for different values of $M$. The solution for $\phi$ is needed to determine the screening factor of the sphere, and thus needed in the region $r < 1$ cm. As can be seen, the solution hardly varies over this region and so we may use the value at $r=0$.

In light of this analysis, we may determine our constraints using Eqs. (S2) and (S6). The range $0.55 < \xi < 0.68$ captures any influence of the chamber geometry, boundary conditions, and parameter variations (disregarding a 1D plane model of the geometry as too unrealistic). The dark blue areas in Fig. 3A-C represent the excluded region with the most conservative value of $\xi=0.55$ while the light blue area indicates the influence of the spread in the values of $\xi$. For the numerical values in the paper, we use the most conservative value of $\xi=0.55$.

These equations predict the acceleration $a$ of atoms from atom-sphere interactions in our vacuum chamber. For experimental convenience, we will express it in units of the earth's acceleration of free fall, $g=(4/3)G\pi r_\oplus \rho_\oplus$, where $r_\oplus$ and $\rho_\oplus$ respectively denote earth's radius and density. Figure **3A** shows our bounds for theories with the exponent $n = 1$. Following one of them from the left (low $M$) to the right, we may distinguish three regions in which the exclusion boundary first points downward (i), then horizontally (ii), and finally upwards (iii). A fourth region (iv) lies above the plotted area and would have the contour pointing vertically upwards.

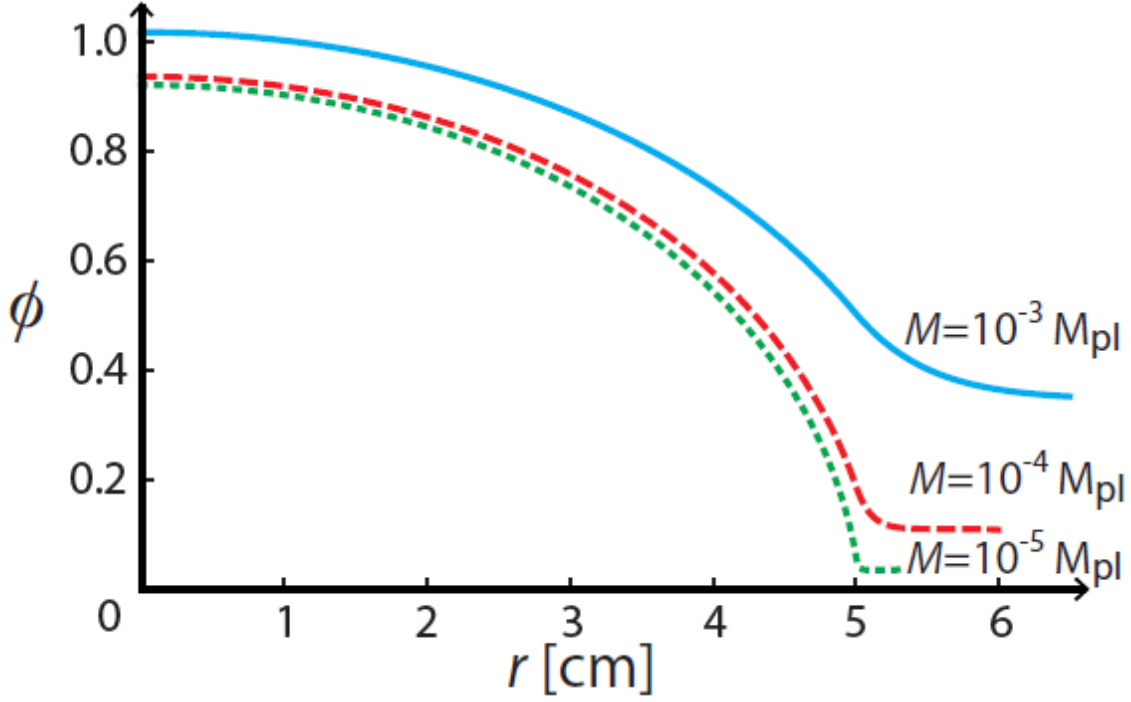

**Fig. S2. Numerical vacuum solution for the scalar field ϕ.** We assume a spherical chamber of radius $r_v = 5$ cm, $n = 1$, and three different values of $M$. The scalar field has been normalized such that $\phi_{vac}$, given by Eq. (S6) with $\xi = 0.6$, corresponds to $\phi = 1$. The plot assumes $\Lambda = 0.1$ meV, close to our lowest bound for $\Lambda$. Simulations with $\Lambda$ between 0.1 and 1 meV and $n = 1$-5 give similar results with $\xi$ not deviating from 0.6 by more than 10%. We see that the field is approximately constant within the region $r$ ~1 cm, where the experiment takes place. Moreover, this constant value agrees well with (S6) with $\xi = 0.6$.

In region (i), both the source and the test mass are screened and the chameleon field reaches its equilibrium value in the vacuum chamber. In this region, the acceleration can be expressed as (neglecting the sphere's gravity)

$$\frac{a}{g} \approx \frac{18 M_{Pl}^2}{r_s \rho_a r_a^2 \rho_\oplus r_\oplus} \left( \frac{nM\Lambda^{4+n}}{\rho_v} \right)^{\frac{2}{1+n}} \frac{r_s^2}{r^2}. \tag{S8}$$

Close to the sphere surface, where $r_s^2/r^2 \sim 1$, this acceleration is large for a small source radius, atom mass, and low vacuum pressure. In the next regime (ii), both the test and source masses are shielded. The chameleon force

$$\frac{a}{g} \approx \frac{18 M_{\text{Pl}}^2}{r_s \rho_a r_a^2 \rho_\oplus r_\oplus} \xi^2 [n(n+1)\Lambda^{4+n} r_v^2]^{2/(n+2)} \frac{r_s^2}{r^2} \tag{S9}$$

is large for a large vacuum chamber and small source radius and atom mass. Going to region (iii) in the direction of higher $M$, the atoms become unscreened but the source mass is still screened. Here the acceleration

$$\frac{a}{g} \approx \frac{6 M_{\text{Pl}}^2}{M r_s \rho_\oplus r_\oplus} \xi [n(n+1)\Lambda^{4+n} r_v^2]^{1/(n+2)} \frac{r_s^2}{r^2} \tag{S10}$$

can be increased by a large vacuum chamber radius and small source radius but is independent of the atom mass. Finally, in region (iv), neither the sphere nor the atoms are screened and the acceleration of the test particle is

$$\frac{a}{g} \approx \frac{2 M_{\text{Pl}}^2}{M^2} \frac{\rho_s}{r_\oplus \rho_\oplus} \frac{r_s^2}{r^2}, \tag{S11}$$

independent of $\Lambda$. In this region, the sensitivity can be increased linearly by increasing the density or the radius of the source object.

Setup

The chameleon force's counter-intuitive behavior informs the design of our setup. A small atomic mass is helpful in regions (i, ii), but the relatively large mass of cesium atoms helps restricting the spatial extent of the atom interferometer by lowering the recoil velocity. This helps us to operate close to the source sphere. A small source radius helps in regions (i-iii), with 1 cm being a good compromise with mechanical requirements.

Our setup has been described in (*13*); we only give the essential details here. After sub-Doppler cooling and optical pumping into the magnetically insensitive $F=4$, $m_F=0$ quantum state, described in the main text, we purify the $m_F=0$ state by two state-sensitive Raman transitions and select a velocity subgroup by a 12 µs, velocity-sensitive Raman pulse.

The cavity consists of a piezo-driven, flat gold mirror and one dielectric mirror having 5-m radius of curvature. The fundamental longitudinal mode of the cavity has a beam waist of 600 µm (located at the surface of the flat mirror), a finesse of $F=100$, and a linewidth of 3.6 MHz. The transverse modes of the cavity are nondegenerate in resonance frequency. The length of the cavity (40.756 cm) sets the free spectral range such that two frequencies separated by the cesium hyperfine splitting of ∼ 9.2 GHz can be simultaneously near-resonant.

The frequency pair is generated from a single laser with low phase noise by a fiber-coupled broadband electro-optic modulator (Eospace). All lasers are diode lasers and frequency stabilized (locked) to a reference laser, which is in turn stabilized to a cesium transition by modulation transfer spectroscopy. The cavity length is stabilized to a tracer laser whose wavelength of 780 nm is far from any transition in cesium and has a negligible effect on the atoms.

We ramp the difference frequency in the Raman frequency pair at a rate of $r\sim 2\pi\times 23$ MHz/s so that the beams remain resonant as the freely falling atoms accelerate. Our normalized detection works by pushing atoms in $F=4$ to the side with our clearing beams, leaving atoms in $F=3$ behind, and then using fluorescence detection of both populations with a camera.

Systematic effects

Systematic effects that are independent of the sphere position are cancelled out in our experiment. This suppresses many systematics typical in atom interferometers to negligible levels, e.g., the Gouy phase and wavefront curvature, laser frequency variations, gravity and gravity gradients, atom density- and index of refraction effects. Table S4 gives an overview of the remaining systematic corrections and errors.

**Table S4.** Corrections for systematic errors and their 1-sigma uncertainties.

| Quantity | Correction [µm/s²] | Uncertainty [µm/s²] |
|---|---|---|
| Magnetic fields | -4.5 | 1.7 |
| AC Stark effect | 1.1 | 0.50 |
| Surface voltage | - | 0.08 |
| Total | -3.4 | 2.1 |

A change in magnetic field can lead to a systematic shift due to the quadratic Zeeman shift of 0.43 kHz/G² of the hyperfine splitting of cesium in the $m_F = 0$ ground state. The symmetry of the Mach-Zehnder interferometer makes a shift from a constant magnetic field common to both interferometer arms. However, a magnetic field gradient $B'$ on top of a constant bias field $B_0$ can add an effective force on the atoms that is proportional to $B_0 B'$ and thus linear in the bias field. This force may be different for measurements with the sphere in the near and far locations due to small changes the MOT position induced, e.g., by slight partial blocking of the MOT beams. To characterize the shift, we run the interferometer at bias fields $B_0$ up to ten times the value of 133 mG used for data taking. Changing the bias field allows us to modulate the effect of any gradient $B'$. Figure S3 shows the change in measured differential acceleration relative to the reference point at 133 mG. Extrapolating to zero field, we find a systematic shift of $(-4.5 \pm 1.7)$ µm/s₂. We note that the effect of eddy currents induced in the sphere by the switching of our MOT magnets is included in this control experiment, because the timing and therefore the magnitude of eddy currents are the same in the control.

The ac Stark effect is an energy-level shift of the atoms induced by the Raman laser pulses (*30*). While wavevector reversal eliminates this effect to leading order, for large changes in the ac Stark shift between pulses, a small second-order influence may remain. To characterize this remaining influence, we increase it by varying the laser intensity by ±20%, much more than the routine intensity changes during our measurement. Figure S4 shows the changes in the acceleration measurement along with their one-sigma confidence interval. Near the nominal laser power $P_0$ without sphere, the measured acceleration is nearly independent of laser power variations. For large deviations of the power, however, a quadratic dependence is evident.

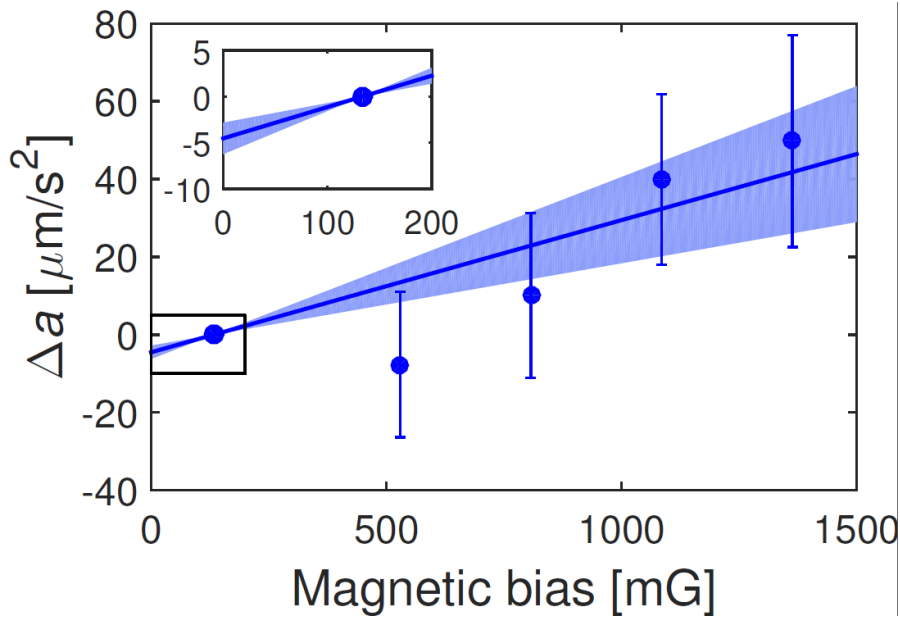

**Fig. S3**. Measurement of the differential acceleration between sphere in the near and far positions versus magnetic bias field. The shaded area is the one-sigma confidence interval (nonsimultaneous function prediction interval) determined from statistics over 1176 experimental runs (7 fringes per point per wavevector direction and sphere position). The inset shows a magnified region near 0 G.

If the sphere has no effect on the interferometry laser pulses, the ac Stark shift systematic will cancel when comparing measurements with the sphere in the near and far positions. Experimentally, however, we observe that placing the sphere in the optical cavity reduces the laser intensity by $4 \pm 1\%$. From Fig. S4, such a reduction in sphere coupling results in a systematic correction of $(1.1 \pm 0.5)$ µm/s$^2$. We note that possible stray light scattered from the atom interferometer laser pulses at the sphere, being 60 GHz off-resonant, would act through an ac Stark effect. It is thus taken out by this control experiment.

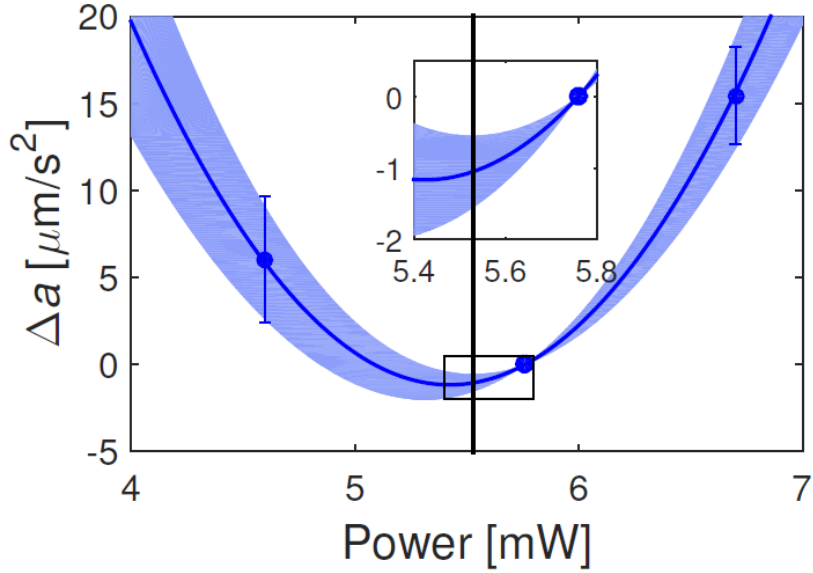

**Fig. S4**. Changes $\Delta a$ in the measured acceleration relative to the acceleration measured at a nominal power of $P_0 \sim 5.8$ mW with the sphere in the far position. The shaded area is the one-sigma confidence interval determined from 12096 experimental runs. The fitting model is $\alpha(P-P_0)(P-\beta)$, where $\alpha$ and $\beta$ are free parameters.

While the sphere is grounded electrically, aluminum surfaces may carry surface potentials because of the approximately 4-10 nm thick insulating natural passivation layer. Thin films of alumina may have a dielectric strength of up to several MV/cm, allowing for surface voltages up to $\sim 10$ V. From the ground state dc polarizability of cesium, even a surface potential of 100 V would cause an acceleration of no more than 0.08 μm/s$^2$ towards the sphere for atoms 5 mm from the surface. The acceleration increases quadratically with voltage. We conservatively use 0.08 μm/s$^2$ as an error bar on the measured acceleration.